\def\scp#1#2{\langle{#1}\vert{#2}\rangle}
\def\bra#1{\langle{#1}\vert}
\def\ket#1{\vert{#1}\rangle}
\def\ah{{\hat{a}}}

\def\ad{{\hat{a}}^\dagger}
\def\bd{{\hat{b}}^\dagger}  
\def\H{{\hat H}}

\def\p{{\hat p}}
\def\q{{\hat q}}

\def\A{{\hat A}}

\documentstyle[eqsecnum,aps]{revtex}

\begin{document}
\draft
\preprint{dias-stp-99-08}
\title{
The principle of symmetric bracket invariance\\ 
as the origin of first and second quantization\footnote{Contributed paper 
to the XIX International Symposium on Lepton and Photon
Interactions at High Energies, Stanford University, August 9-14, 1999.}\\}
\author{T.~Garavaglia \cite{email}\cite{byline}\\}
\address{Institi\'uid \'Ard-l\'einn Bhaile \'Atha Cliath, 
Baile \'Atha Cliath  4, \'Eire\\}
\author{S.~K.~Kauffmann}
\address{
Unit 3, 51-53 Darley Street, Mona Vale, NSW 2103, Australia\\
}
\date{8 July, 1999}
\maketitle
\begin{abstract}
\par            
The principle of invariance of the c-number symmetric bracket is used to derive 
both the quantum operator commutator relation $[\hat q, \hat p]=i\hbar$ and the 
time-dependent Schr\"odinger equation. A c-number dynamical equation is 
found
which leads to the second quantized field theory of bosons and 
fermions. 
\end{abstract}
\centerline{dias-stp-99-08 also hep-th/9907059}
\widetext
\section{Introduction:\protect\\ 
}
\par
Occasionally in the development of quantum theory and quantum field theory, 
something fundamental and simple is overlooked. This is the case with the 
introduction of the ordered Poisson bracket and its consequences. It is shown 
in this paper that the time-dependent Schr\"odinger equation and the 
commutation relation between position and momentum, the quantum bracket $[\hat 
q, \hat p]=i\hbar$ \cite{Dirac}, is in fact a consequence of the principle of 
invariance under a one parameter canonical transformation of the c-number 
symmetric bracket. Furthermore, the relation between expectation values and 
classical dynamics, and the probability interpretation of quantum theory are a 
consequence of this procedure. In addition, a c-number dynamical equation is 
derived, which provides the fundamental condition for the boson and fermion 
operator commutation relations. 
\par 
Although the idea of the symmetric analog of the Poisson bracket has 
appeared in the theory of differential geometry and algebraic ideals 
\cite{D-V}, and in classical constraint dynamics \cite{FandK}, its clear relevance to 
fundamental physics has not until now been demonstrated. The idea of the 
ordered Poisson bracket and related symmetric and antisymmetric brackets has 
been introduced in \cite{GandK} to provide a c-number analog of the usual 
boson commutator and fermion anticommutator for quantum fields. From the basic 
concept of the ordered bracket, the antisymmetric and symmetric brackets are 
defined. The principle of invariance of the antisymmetric bracket under a one 
parameter canonical transformation leads to Hamilton's dynamical equations, 
and the generator of this transformation is the Hamiltonian. What is new and 
surprising is that the analogous property for the symmetric bracket leads to 
Schr\"odinger's equation, and the generator of the one parameter canonical 
transformation in this case is the expectation value of the Hamiltonian 
operator. Furthermore; these c-number brackets provide a natural derivation of 
the boson and fermion commutation relations, when operator infinitesimal time 
development equations are sought which have the c-number equations as a 
displacement state expectation value. 
\par 
In this paper dimensionless phase space coordinates are used such that 
$q_i\rightarrow q_i/q_o$, $p_i\rightarrow p_i/p_o$, and $q_o p_o=\hbar$. For a 
given mass $M_o$, the natural units of length, time, and energy are 
respectively, 
 $\lambda={2\pi{\hbar}/{M_oc}}$, $T_o={2\pi{\hbar}/{M_oc^2}}$, and 
${E_o}=M_oc^2=\hbar\omega_o$. If $M_o$ is chosen to be the Planck mass,
$M_P=\sqrt{\hbar c/G_N}$, then these units can be expressed in terms of natural physical
constants(Planck's reduced constant $\hbar$, the 
speed of light $c$, and the 
Newtonian gravitational constant $G_N$).

\section{Complex phase space and c-number brackets}
\par
Ordinary classical dynamics is usually discussed in terms
of real-valued phase space vector variables of the form
$(\vec q,\vec p)$.  However, its relation to quantum
theory and to fermion systems is much more transparent if
one changes these real phase space vector variables to the
complex-valued dimensionless phase space vector variables
$\vec a \equiv (\vec q + i\vec p/)/\sqrt{2}$
and their complex conjugates
$\vec a^{\, *} = (\vec q - i\vec p)/\sqrt{2}$.
In terms of components of both of these types
of phase space vector variables, the usual Poisson bracket of
ordinary classical dynamics is
\begin{eqnarray}
\{f,\, g\} &&\equiv \sum_k\left ({\partial f\over\partial q_k}
{\partial g\over\partial p_k} - {\partial g\over\partial q_k}
{\partial f\over\partial p_k}\right )\nonumber\\
&&= -\,{i}
\sum_k\left ({\partial f\over\partial a_k}
{\partial g\over\partial a^*_k} - {\partial g\over\partial a_k}
{\partial f\over\partial a^*_k}\right ).
\label{notes1}\end{eqnarray}
From the second Poisson bracket representation given
above, we abstract the semi-bracket, 
which we call the ordered Poisson bracket,
\begin{equation}
\{f\circ g\}=\sum_k{\partial f\over\partial a_k}
{\partial g\over\partial a^*_k}
={\partial f\over\partial{\vec a}}\cdot
{\partial g\over\partial{\vec a}^*}.
\label{notes2}\end{equation}
We note
that while $\{f\circ g\}$ is linear in each of its two argument
functions $f$ and $g$, it is neither antisymmetric nor symmetric
under their interchange.  However, it does
satisfy the identity $\{f\circ g\} = \{g^*\circ f^*\}^*$, which
is in algebraic correspondence with the Hermitian conjugation
formula for the product of two Hilbert-space operators,
i.e., $\hat f\hat g = (\hat g^{\dagger}\hat f^{\dagger}
)^{\dagger}$.  
From Eq. (\ref{notes2}) we define the c-number symmetric and antisymmetric brackets
\begin{equation}
\{f,\, g\}_{\pm} \equiv \{f\circ g\} \pm \{g\circ f\},
\label{{notes2a}}\end{equation}
where we note $\{f,g\}=-i\{f,g\}_-$.
We readily calculate the c-number symmetric and antisymmetric 
brackets for the components of $\vec a$ and $\vec a^{\, *}$,
\begin{equation}
\{a_i,\, a_j\}_{\pm} = 0 = \{a_i^*,\, a_j^*\}_{\pm}, \quad
\{a_i,\, a_j^*\}_{\pm} = \delta_{ij} = \pm\{a_j^*,\, a_i\}_{\pm}.
\label{notes4}\end{equation}
\par
Infinitesimal canonical transformations, which leave the brackets invariant, 
are now introduced.
The canonical transformations of ordinary classical dynamics
are mappings of the complex phase space vectors $\vec a\to\vec
A(\vec a,\vec a^{\, *})$ and
$\vec a^{\, *}\to \vec A^*(\vec a,\vec a^{\, *})$,
which preserve the antisymmetric
c-number Poisson bracket relations among the complex phase space vector
components. Also we consider the canonical transformations 
of complex vector phase space mappings which preserve the c-number symmetric  
bracket relations among the complex phase space vector components. It is 
important to note that the complex phase space vectors are related to ordinary 
classical mechanics phase space coordinates in the case of the antisymmetric 
bracket; however, in the symmetric bracket case they correspond to the 
expansion coefficients of either quantum wave functions or the c-number
limit of quantum fields.                
\par
Specializing now to infinitesimal phase space transformations
$\vec a\to\vec A = \vec a + \delta\vec a(\vec a,\vec a^{\, *})$,
we readily calculate the c-number antisymmetric and symmetric 
brackets for the components of $\vec A$ and $\vec A^*$ to first
order in $\delta\vec a$ and $\delta\vec a^{\, *}$,
\begin{equation}
\{A_i,\, A_j\}_{\pm} =
{\partial (\delta a_j)\over\partial a_i^*}\pm
{\partial (\delta a_i)\over\partial a_j^*}\, , \quad
\{A_i^*,\, A_j^*\}_{\pm} =
{\partial (\delta a_i^*)\over\partial a_j}\pm
{\partial (\delta a_j^*)\over\partial a_i}\, ,
\label{notes6}\end{equation}
\begin{equation}
\{A_i,\, A_j^*\}_{\pm} = \delta_{ij} +
{\partial (\delta a_i)\over\partial a_j} +
{\partial (\delta a_j^*)\over\partial a_i^*} =
\pm\{A_j^*,\, A_i\}_{\pm}.
\label{notes7}\end{equation}
If we now impose the requirement that this infinitesimal phase
space vector transformation 
preserves the c-number antisymmetric or symmetric bracket
relations among the complex phase space vectors,
we obtain the three equations,
\begin{equation}
{\partial (\delta a_j)\over\partial a_i^*} = \mp
{\partial (\delta a_i)\over\partial a_j^*}\, , \quad
{\partial (\delta a_j^*)\over\partial a_i}= \mp
{\partial (\delta a_i^*)\over\partial a_j}\, , \quad
{\partial (\delta a_i)\over\partial a_j} +
{\partial (\delta a_j^*)\over\partial a_i^*} = 0.
\label{notes8}
\end{equation}
The last of these equations is independent of the value of
the $\mp$ symbol, and it is satisfied in
particular for one-parameter infinitesimal $\delta\vec a$ which
are of the form
\begin{equation}
\delta a_i = -\,{i}(\delta\lambda)
{\partial G\over\partial a_i^*},\qquad \delta a_j^* = {i}(\delta\lambda)
{\partial G\over\partial a_j},
\label{notes9}\end{equation}
where $\delta\lambda$ is a real-valued infinitesimal parameter
and $G(\vec a,\vec a^{\, *},\lambda)$ is a real-valued generating function.
We thus can readily verify that the last 
equation in Eq. (\ref{notes8}) is satisfied.
From the two immediately preceding equations Eq. (\ref{notes9}), we
obtain the form of the equation which governs any continuous
one-parameter trajectory of sequential infinitesimal canonical
transformations in the complex vector phase space:
\begin{equation}
i{da_i\over d\lambda} =
{\partial G\over\partial a_i^*} \quad \hbox{or} \quad
-i{da_i^*\over d\lambda} =
{\partial G\over\partial a_i}\, .
\label{notes11}\end{equation}
In the most general circumstance, $G$ may have an explicit
dependence on $\lambda$. 
These equations may be rewritten as the pair of real equations:
\begin{equation}
{dq_i\over d\lambda} =
{\partial G\over\partial p_i}\, , \quad
{dp_i\over d\lambda} =
-\, {\partial G\over\partial q_i}\, ,
\label{notes12}\end{equation}
which are generalized Hamilton's equations.

For the case of ordinary classical dynamics, antisymmetric bracket case 
 (for which  $\mp = +$ in Eq. (\ref{notes8})), the first two of the group of three
equations which were given above are satisfied identically
for the one-parameter infinitesimal $\delta\vec a$ of the
generating function form which has just been given in Eq. (\ref{notes9}).
However, for the symmetric bracket case 
(for which  $\mp = -$ in Eq. (\ref{notes8})), the first two of
that group of three equations impose the following constraint on
those real-valued generating functions
$G(\vec a,\vec a^{\, *},\lambda)$ of continuous one-parameter
canonical transformation trajectories:
\begin{equation}
{\partial^2 G\over\partial a_i\partial a_j} = 0 =
{\partial^2 G\over\partial a_i^*\partial a_j^*}\, .
\label{notes13}\end{equation}

\section {Fermion c-number dynamics}
\par
For the symmetric bracket case, which we call fermion c-number dynamics, the
generating functions of the continuous
one-parameter trajectories of sequential infinitesimal
canonical transformations 
are constrained to be constant or linear in
each of $\vec a$ and $\vec a^{\, *}$, as well as real-valued.
The most general form for the  
generating function is therefore
\begin{equation}
G(\vec a,\vec a^{\, *},\lambda) = G_0(\lambda) +
\sum_k\left (\tilde g_k(\lambda)a_k^* + \tilde g_k^*(\lambda)a_k\right ) +
\sum_l\sum_mG_{lm}(\lambda)a_l^*a_m,
\label{notes14}\end{equation}
where $G_0(\lambda)$ is real and $G_{lm}(\lambda)$ is a
Hermitian matrix.  Upon putting this constrained form
for $G$ into the complex phase space form of the
generalized Hamilton's equations Eq. (\ref{notes11}),
we arrive at
\begin{equation}
i{da_i\over d\lambda} =
\tilde g_i(\lambda) + \sum_j G_{ij}(\lambda)a_j,
\label{notes15}\end{equation}
which is a (possibly) inhomogeneous linear equation of matrix
Schr\"odinger form.
If $\tilde g_i(\lambda) = 0$, the preceding equation is a
general homogeneous type of Schr\"odinger equation, whereas
if
$\tilde g_i(\lambda) \propto\delta(\lambda - \lambda')$,
it is a general propagator type of Schr\"odinger equation.
It is clear that the c-number dynamics of the symmetric bracket case
must be linear and describable by a Schr\"odinger type equation.
\par
The generating functions of the continuous one-parameter
canonical transformation trajectories are usually considered
to be observables of classical theory when they
have no explicit dependence on the parameter.  
Thus we restrict $G(\vec a,\vec a^{\, *},\lambda)$ to have  
no  explicit
$\lambda$-dependence. 
In the present case it is
always possible to suppress the inhomogeneous
part if the Hermitian matrix $G_{lm}$ is
not singular.  This is done by making the canonical
transformation
\begin{equation}
a_i\to A_i = a_i + \sum_j\left (G^{-1}\right )_{ij}\tilde g_j.
\label{notes16}\end{equation}
It is easily verified that these transformed $A_i$
also satisfy the c-number symmetric
bracket relations.
In terms of these $A_i$'s, the generalized Hamilton's
equations become
\begin{equation}
i{dA_i\over d\lambda} =
\sum_j G_{ij}A_j,
\label{notes17}\end{equation}
which are of the homogeneous Schr\"odinger matrix equation form,
while $G$ itself becomes
\begin{equation}
G(\vec A,\vec A^*) = G_0 -
\sum_l\sum_m\left (G^{-1}\right )_{lm}\tilde g_l^*\tilde g_m +
\sum_l\sum_mG_{lm}A_l^*A_m,
\label{notes18}\end{equation}
which has no inhomogeneous term.
\section{Derivation of the time-dependent Schr\"odinger equation}

\par 
The result found in Eq. (\ref{notes17}) following from the invariance of the 
symmetric bracket can now be used to derive the time-dependent Schr\"odinger 
equation. Choosing the parameter $\lambda$ to be a time parameter $t$ and 
assuming that the canonical transformation Eq. (\ref{notes16}) has been made, the 
dynamical equation Eq. (\ref{notes17}) for a time-independent $G_{ij}$ becomes 
\begin{equation}
i{\dot a}_i(t)=\{g(\vec a,\vec a^{\, *}),\, a_i(t)\}_{+}=\sum_jG_{ij}a_j(t).
\label{se1}\end{equation}
Keeping the last term only in Eq. (\ref{notes18}) 
and changing 
$\vec 
A\rightarrow 
\vec a$ and 
$G(\vec A,\vec A^{\, *})\rightarrow g(\vec a,\vec a^{\, *})$,
the real valued generating function becomes 
\begin{equation}
g(\vec a,\vec a^{\, *})=\sum_i\sum_ja^*_i(t)G_{ij}a_j(t).
\label{se2}\end{equation}
The Hermitian matrix element $G_{ij}$ is associated with an Hermitian 
operator $\hat G$ such that
\[ G_{ij}=\bra{i}{\hat G}\ket{j},\]
where ${\ket{i}}$ form a orthonormal complete set of states with identity
operator $I=\sum_i \ket{i}\bra{i}$.
A general state expanded in this basis is
\begin{equation}
\ket{\psi(t)}=\sum_i a_i(t)\ket{i},
\label{se3}\end{equation}
with $a_i(t)=\scp{i}{\psi(t)}$.
From Eq. (\ref{se1}), 
follows the 
relation
\begin{eqnarray}
\sum_i{\dot a}_i(t)\ket{i} 
&&=i{\partial \over \partial t}\ket{\psi(t)}=\sum_i\sum_j\ket{i}G_{ij}a_j(t)\nonumber \\
&&=\sum_i\sum_j\ket{i}\bra{i}{\hat G}\ket{j}a_j(t)\nonumber \\
&&=\sum_i\sum_j\ket{i}\bra{i}{\hat G}\ket{j}\scp{j}{\psi(t)}=I{\hat G}\ket{\psi(t)}\nonumber \\
&&i{\partial \over \partial t}\ket{\psi(t)}={\hat G}\ket{\psi(t)},\nonumber \\
\label{se4}
\end{eqnarray}
which is the time-dependent Schr\"odinger equation when $\hat G$ is identified 
with the Hamiltonian operator $\hat H(\q,\p)$.  
\par
We see here the difference in the interpretation of the quantities $a_i(t)$ in 
the case of the antisymmetric and symmetric brackets. In the former these are 
just the complex coordinates associated with position $q_i$ and momentum 
$p_i$; whereas, in the latter, they represent the expansion coefficients of a 
general quantum state in terms of an orthonormal basis. Both brackets lead to 
the completeness relation 
\begin{equation}
\bra{q}\{\ket{\psi(t)},\bra{\psi(t)}\}_{\pm}\ket{q'}=\delta(q-q').
\label{se5}\end{equation}
This is seen from
\begin{eqnarray}
&&\bra{q}\{\ket{\psi(t)},\bra{\psi(t)}\}_{\pm}\ket{q'}\nonumber \\
&&=\sum_i\sum_j\bra{q}\{a_i,a^*_j\}_{\pm}\ket{i}\bra{j}{q'}\rangle\nonumber \\
&&=\sum_i\sum_j\bra{q}\delta_{ij}\ket{i}\bra{j}{q'}\rangle=\bra{q}I\ket{q'}=\delta(q-q').\nonumber \\
\label{se6}
\end{eqnarray}

\section{Derivation of $[\q,\p]=\lowercase{i}\hbar$}

\par 
The principle of symmetric bracket invariance leads to quantum mechanics 
because it leads to the time-dependent Schr\"odinger equation and to a derivation of 
the Dirac bracket relation $[\hat q, \hat p]=i.$
Firstly, one considers the results of the invariance of the antisymmetric
bracket under a one parameter canonical transformation. 
The time development of a real function $f(\vec a,\vec a^*,t)$ is given by
\begin{equation}
\dot f=-i\{f,H\}_{-}
+{{\partial  f}\over{\partial  t}},
\label{(1)}\end{equation}
and the dynamical equations for the coordinates are
\begin{eqnarray}
\dot a_i &&=-i\{a_i,H\}_{-}=-i{{\partial  H}\over{\partial  a_i^*}}\nonumber \\
\dot a_i^* &&=-i\{a_i^*,H\}_{-}=i{{\partial  H}\over{\partial  a_i}}.\nonumber \\
\label{(2)}\end{eqnarray}
These are equivalent to Hamilton's equations of classical mechanics.
\par
The invariance of the symmetric bracket under a one parameter canonical 
transformation gives dynamical equations for the coordinates(wave function 
expansion coefficients in this case). It is convenient to write Eq. (\ref{se1}) and 
its complex conjugate as 
\begin{eqnarray}
i{{\partial  {\vec a}}\over{\partial  t}}&&=\hat G\cdot {\vec a}\nonumber \\
-i{{\partial  {\vec a^*}}\over{\partial  t}}&&={\vec a^*}\cdot\hat G.\nonumber \\
\label{(3)}\end{eqnarray}
These are Schr\"odinger's equations for $a_i$ and $a_i^*$ (i can also be a 
continuous index). The time development of a real 
function $\bar f(\vec a,\vec a^*,t) 
={\vec a^*}\cdot\hat F \cdot {\vec a}$, which depends on 
the generator for the one parameter canonical transformation
$g(\vec a,\vec a^*)={\vec a^*}\cdot\hat G \cdot {\vec a}$, is given by
\begin{eqnarray}
{\dot{\bar f}}&&={\dot{\vec a^*}}\cdot\hat F \cdot {\vec a}+
{\vec a^*}\cdot\hat F \cdot {\dot{\vec a}}+
{\vec a^*}\cdot{\partial\hat F\over \partial t} \cdot {\vec a}\nonumber \\
{\dot{\bar f}}&&=-i {\vec a^*}\cdot [\hat F,\hat G]\cdot{\vec a}
+{\overline{\partial  f}\over{\partial  t}},\nonumber \\
\label{(4)}\end{eqnarray}
which follows from Eq. (\ref{(3)}).
Here $\hat F$ and $\hat G$ are Hermitian matrices(operators). 
For the discrete index case
\begin{equation}
\bar f(\vec a,\vec a^*,t)         
=\sum_i\sum_j\scp{\psi(t)}{i}\bra{i}\hat F(t)\ket{j}\scp{j}{\psi(t)}  
=\bra{\psi(t)}\hat F(t)\ket{\psi(t)},
\label{(4a)}\end{equation}
and  
\begin{equation}
\bar f(\vec a,\vec a^*,t)
=\int\int\scp{\psi(t)}{p}\bra{p} \hat F(t) \ket{p'}\scp{p'}{\psi(t)} dpdp',
\label{(4b)}\end{equation}
for the continuous index $p$. The form of $g({\vec a},{\vec a}^*)$ shows that 
classical results are to be associated with  expectation values. When $\hat G$ 
is identified with the Hamiltonian operator $\hat H$ and $\ket{i}$ is an 
eigenstate of the Hamiltonian with eigenvalues $E_i$, the bilinear form of 
$g({\vec a},{\vec a}^*)$ leads to the statistical interpretation of quantum 
mechanics. This is seen from 
\begin{equation}
\bar H=\vec a \cdot\hat H\cdot \vec a
=\sum_i E_i\vert a_i\vert^2=\bra{\psi(t)}\hat H\ket{\psi(t)},
\label{(4c)}\end{equation}
with 
\[ \scp{\psi}{\psi}=\sum_i \vert a_i\vert^2=1. \]
\par 
The classical dynamics case(antisymmetric bracket result Eq. (\ref{(1)}) 
for the Hamiltonian 
\begin{equation}
H(a,a^*)={{p^2}\over{2m}}+V(q)
\label{(5)}\end{equation} 
gives the result for $f(a,a^*)=q$ that
$\dot q=p/m$. For the c-number symmetric bracket result Eq. (\ref{(4)}) 
to give a result for $\bar f$ that corresponds to
classical mechanics, one identifies $\hat G$ with the Hamiltonian
operator $\hat H(\p,\q)$, and observes that 
$\dot{\bar q}=\bar p/m$ when $[\hat q,\hat p]=i$. 
This is found from the expectation value
\begin{equation}
\bar f=\bra{\psi(t)}\hat f(t)\ket{\psi(t)}
=\bra{\psi}\hat U^\dagger(t)\hat f(t)\hat U(t)\ket{\psi},
\label{(5a)}\end{equation}
with $\ket{\psi}=\ket{\psi(0)}$, $\hat U(t)=exp(-it\hat H)$ and the 
relation
\begin{equation}
{d\bar f\over dt}=\bra{\psi(t)}i[\hat H,\hat f]
+{\partial \over \partial t}\hat f(t)\ket{\psi(t)},
\label{(5b)}\end{equation}
which corresponds to Eq. (\ref{(4)}) when Eq. (\ref{se3}) is used.
Choosing $\hat f(t)=\hat q$ and using the Hamiltonian operator found from
Eq. (\ref{(5)}), one finds
\begin{equation}
{d\bar q \over dt}=\bra{\psi(t)}i[{\p^2\over 2m},\hat q]\ket{\psi(t)}
={\bar p\over m}
\label{(5c)}\end{equation}
when $[\hat q,\hat p]=i$, which of course is equivalent to $[\ah,\ad]=1$.
The appropriate correspondence between force and the potential function
follows from Eq. (\ref{(5b)}) when $\hat f(t)=\p$. This gives the result
\begin{equation}
{d\bar p\over dt}=-\bra{\psi(t)}{\partial V(q)\over \partial q}\ket{\psi(t)},
\label{(5d)}\end{equation}
since the quantum bracket relation between $\p$ and $\q$ implies
\begin{equation}
[\p,V(q)]=-i{\partial V(q)\over \partial q}.
\label{(5e)}\end{equation}
This is of course the well known result of Ehrenfest \cite{ehren}, and
the appropriate states to use in the evaluation of these 
expressions when associating them with corresponding classical equations is
the minimum uncertainty displacement states discussed in a Sec. VI.
These results  clearly shows that  
quantum mechanics is a consequence of the principle of symmetric bracket
invariance, and this is
clearly an advance in the understanding of the origin and properties 
of quantum theory.
\par 
The above proof depends upon the association of the quantum operators with 
observed quantities through the prediction of distributions for the spectrum 
of the operators and expectation values. Naturally, the generating function  
$g(\vec a^*,\vec a)$ given in Eq. (\ref{se2}) leading to the invariance of the 
symmetric bracket, is of this form. Since $\hat G$ can be identified with the 
Hamiltonian $\hat H(\q,\p)$, this requires the existence of the expectation 
values for the operators $\hat p$ and $\hat q$. A related derivation of the 
result $[\hat q,\hat p]=i$, which depends upon the association of 
distributions and  expectation values of the selfadjoint operators $\hat q$ 
and $\hat p$ with the classically observed values, is found in \cite{TGPRL}, 
and a similar approach is found in \cite{ah}. The argument of Dirac 
\cite{Dirac} leading to $[\hat q,\hat p]=i$ is incorrect because it depends 
upon the non-classical concept of non-commuting quantities in the definition 
of the classical Poisson bracket. 
\par
It is easily seen that the argument above for the non-relativistic Hamiltonian 
leading to the quantum bracket result applies to the relativistic Dirac 
Hamiltonian associated with a fermion. Furthermore, the importance of the 
natural relation between the expectation value of an operator and its observed 
classical values, which emerges from the principle of invariance of the 
symmetric bracket, also resolves the dilemma of Dirac where he finds the 
eigenvalues of $\dot {\hat q}$ to be $\pm c$, \cite{Dirac2}.  The correct 
result for a free relativistic Dirac particle of mass $m$, momentum $p$, and 
energy $E$ is 
\begin{equation}
\dot{\bar q}={p\over E}=\beta,
\label{(6)}\end{equation}
where, using $\hbar=c=1$ and the conventions of \cite{LandL},
\begin{eqnarray}
p&&=\gamma \beta m\nonumber \\
E&&=\gamma m\nonumber \\
\gamma&&=1/\sqrt{1-\beta^2}.\nonumber \\
\label{(7)}\end{eqnarray}
This follows from the time derivative of the expectation value
\begin{eqnarray}
\dot{\bar q}&&={d\over dt}\bra{\psi(t)}\hat q\ket{\psi(t)}\nonumber \\
&&=-i\bra{\psi(t)}[\hat q,\hat H]\ket{\psi(t)}\nonumber \\
&&=\bar u(p)\vec{\gamma}u(p)/2E={\vec{p}\over E},\nonumber \\
\label{(8)}\end{eqnarray} 
where for a free Dirac particle
\begin{eqnarray}
\hat H &&=\gamma^0\vec\gamma\cdot \vec p + \gamma^0 m,\nonumber \\
\scp{x}{\psi(t)} &&=\psi(x,t)={1\over \sqrt{2E}}u(p)e^{-ipx}\nonumber \\
px &&=p^0t-\vec p\cdot \vec r\nonumber \\
\bar u(p) &&=u^\dagger(p)\gamma^0\nonumber \\
\bar u(p)u(p) &&=2m.\nonumber \\
\label{(9)}\end{eqnarray}
This removes the need for the notion of zitterbewegung, which is associated 
with the Heisenberg operator but not with the observed mean value of the 
operator through the expectation value. 

\section{Quantum field operators associated with $\lowercase{a_i}$ 
and $\lowercase{a_i^*}$}

\par
For each index $i$, one can associate an operator with the complex numbers
$a_i$ through the matrix element
\begin{equation}
a_i=\bra{a_i}\ah_i\ket{a_i}.
\label{o1}\end{equation}
As shown in the next section,
the operator relations which are consistent with the infinitesimal 
time development 
equations for both bracket relations Eq. (\ref{se1}) and Eq. (\ref{(2)}) 
involving the complex numbers $a_i$ are
\begin{equation}
\{a_i,\, a_j^*\}_{\pm} = \delta_{ij}=[\ah_i,\ad_j]_{\pm}.
\label{o2}\end{equation}
Introducing the notation $a=a_i$, we can discuss both the case of boson 
operators and fermion operators without loss of generality.
In both cases, the states to use in Eq. (\ref{o1}) are defined as 
displacement states
\begin{equation}
\ket{a}={\hat D}(a)\ket{0}.
\label{o3}\end{equation}
In the boson case, the displacement operator is
\begin{equation}
{\hat D}(a)=e^{a\ad-a^*\ah},
\label{o4}\end{equation}    
such that
\begin{eqnarray}
\sigma(q) &&=\sigma(p)=1/\sqrt{2},\quad \sigma(q)\sigma(p)=1/2 \nonumber \\
\sigma^2(A) &&=\bra{a}\A^2\ket{a}-\bra{a}\A\ket{a}^2\nonumber \\
a &&=\bra{a}\ah\ket{a},\quad a^*=\bra{a}\ad\ket{a}.\nonumber \\  
\label{o5}\end{eqnarray}
The interpretation of $a_i$ in this case is clear. The state $\ket{a_i}$ is 
the minimum uncertainty state, and the $a_i$'s are the complex numbers that 
appear in the antisymmetric bracket Eq. (\ref{notes4}) and Hamilton's equations, i.e. 
classical coordinates. 
\par
The fermion case can be treated in a similar manner; however, there are some 
modifications in interpretation. The displacement operator in this case is 
\begin{eqnarray}
{\hat D}(\xi) &&=e^{\xi\ad-\xi^*\ah},\quad \xi=|\xi|e^{+i\phi}\nonumber \\
\ket{a} &&={\hat D}(\xi)e^{-i\phi/2}\ket{0}=\cos(|\xi|)e^{-i\phi/2}\ket{0}
+e^{+i\phi/2}\sin(|\xi|)\ket{1}.\nonumber \\
\label{o6}\end{eqnarray}
This give the following:
\begin{eqnarray}
a &&=\bra{a}\ah\ket{a}={\sin2|\xi| \over 2}e^{i\phi}\nonumber \\
a^* &&=\bra{a}\ad\ket{a}={\sin2|\xi| \over 2}e^{-i\phi}\nonumber \\
\bra{a}\ad\ah\ket{a} &&=\sin^2|\xi|,\quad 
\bra{a}\ah\ad\ket{a}=\cos^2|\xi|\nonumber \\
 &&\bra{a}\ah\ad +\ad\ah\ket{a}=1,\nonumber \\
\label{o7}\end{eqnarray}
when
\begin{eqnarray*}
\ah\ket{0} &&=\ad\ket{1}=0 \nonumber \\
\ah\ket{1} &&=\ket{0},\quad\ad\ket{0}=\ket{1}. \nonumber \\
\end{eqnarray*}
An analogous calculation for $\sigma(q)$ and $\sigma(p)$ for the fermion case 
gives 
\begin{eqnarray}
\sigma(q) &&=(1-\sin^2(2|\xi|)\cos^2(\phi))^{1/2}/\sqrt{2}\nonumber \\
\sigma(p) &&=(1-\sin^2(2|\xi|)\sin^2(\phi))^{1/2}/\sqrt{2}\nonumber \\
\sigma(q) &&\sigma(p)\geq 0.\nonumber \\
\label{o8}\end{eqnarray}
The last inequality in Eq. (\ref{o8})  does not violate the minimum uncertainty 
inequality, $\sigma(q)\sigma(p)\geq 1/2$, since $[\q,\p]\neq i$, and $\q$ 
and $\p$ are not conjugate coordinates. 

\section{Infinitesimal c-number transformations and their relation to boson and fermion 
operators}

\par
The infinitesimal transformations induced by the c-number symmetric and 
antisymmetric brackets have analogous relations involving operators, and these 
lead naturally to the boson and fermion operator relations Eq. (\ref{o2}). For the 
antisymmetric bracket, the transformation associated with a time $dt$ is 
\begin{equation}
a_i(dt)=a_i(0)+idt\{H(\vec a,\vec a^*),a_i(0)\}_-,
\label{it1}\end{equation}
and the appropriate operator equation to associate with this c-number equation
is
\begin{equation}
\ah_i(dt)=\ah_i(0)+idt[{\hat H}(\ah,\ad),\ah_i(0)].
\label{it2}\end{equation}
With $\ah_i=\ah_i(0)$, the commutation relation for $\ah_i$ and $\ad_j$ in this case follows from the 
relation $[\q,\p]=i$, which is a consequence of Eq. (\ref{(5c)}), and is the boson 
commutator 
\begin{equation}
[\ah_i,\ad_j]=\delta_{ij}.
\label{it3}\end{equation}
The associated classical Hamiltonian is found from the normal ordered matrix 
element
\begin{equation}                        
H(\vec a,\vec a^*)=\bra{\vec a}:\hat H:\ket{\vec a},
\label{it4}\end{equation}
with $\ket{\vec a}=\ket{ a_0}\ket{ a_1}\cdots\ket{a_n}$,
where $\ket{ a_i}$ are the minimum uncertainty states defined 
in Eq. (\ref{o4}).
Here normal ordering is defined as moving the operators $\ad_i$ to the left
according to the boson commutation operation. In this way, the c-number 
equation Eq. (\ref{it1}) is a consequence of the expectation 
value of Eq. (\ref{it2}),
using the displacement states $\ket{\vec a}$ found from Eq. (\ref{o4}). 
\par                                            
The infinitesimal c-number transformation associated with the symmetric 
bracket implies both the boson commutation and fermion anticommutation 
relations. For the c-number symmetric bracket, the infinitesimal 
transformation is 
\begin{equation}
a_i(dt)=a_i(0)-idt\{g(\vec a,\vec a^*),a_i(0)\}_+,
\label{it5}\end{equation}
and the appropriate operator equation to associate with this is
\begin{equation}
\ah_i(dt)=\ah_i(0)+idt[{\hat g}(\ah,\ad),a_i(0)],
\label{it6}\end{equation}
with 
\begin{equation}
\hat g=\vec \ad\cdot\hat G\cdot \vec \ah,
\label{it7}\end{equation}
and $\ah_i=\ah_i(0)$.
It is now shown that Eq. (\ref{it6}) yields the c-number 
equation Eq. (\ref{it5}) when 
the operators satisfy
either the boson commutator or fermion anticommutator relation Eq. (\ref{o2}).
This follows from using
\begin{eqnarray}
[\hat g,\ah_i] &&=-\sum_j\sum_k([\ah_i,\ad_j]\ah_k+\ad_j[\ah_i,\ah_k])G_{jk} \nonumber \\
 &&=-\sum_j\sum_k(\delta_{ij}\ah_k)G_{jk},\, boson\,\, case \nonumber \\
 &&=-\sum_j\sum_k((1-2\ad_j\ah_i)\ah_k+2\ad_j\ah_i\ah_k)G_{jk},\, fermion\,\, case  \nonumber \\
 &&\equiv -
\sum_k 
G_{ik}\ah_k, \nonumber \\
\label{it9}\end{eqnarray}
and one finds
\begin{eqnarray}
\ah_i(dt) &&=\sum_j (\delta_{ij}-idtG_{ij})\ah_j \nonumber \\
\bra{\vec a}\ah_i(dt)\ket{\vec a}
 &&=\sum_j \bra{\vec a}(\delta_{ij}-idtG_{ij})\ah_j\ket{\vec a} \nonumber \\
a_i(dt) 
&&=\sum_j 
(\delta_{ij}-idtG_{ij})a_j, \nonumber \\
\label{it10}\end{eqnarray}
which agrees with Eq. (\ref{se1}) and Eq. (\ref{it5}).
Here the state $\ket{\vec a}$ is defined as the direct product of 
displacement states,
$\ket{\vec a}=\ket{ a_0}\ket{ a_1}\cdots\ket{a_n}$, found 
from either 
Eq. (\ref{o4}) for the boson case or Eq. (\ref{o6}) for the fermion case.

\section{Quantum fields}

\par It is seen from the above that the infinitesimal transformations obtained 
in both the antisymmetric and symmetric bracket case have corresponding 
operator equations, if the operators $\ah_i$ and $\ad_j$ satisfy the boson 
commutation relations in the former case and the fermion anticommutation 
relations in the latter. Thus the expansion of quantum fields in these 
operators is a natural consequence of the relations found for the associated 
c-numbers. In both cases the usual quantum field expansion \cite{LandL} is 
\begin{equation}
\Psi(\vec r,t)=\sum_i(\ah_i(t)\psi_i(\vec r)+\bd_i(t)\psi_i^*(\vec r)),
\label{qf1}\end{equation}
where $\bd_i(t)=\ah_i(p_{i0}<0)$, with four-momentum time 
component $p_0$, 
is an antiparticle creation operator.
The associated c-number fields are found by forming the matrix 
element with the displacement state $\ket{\vec a}$ appropriate to either 
the boson or the fermion case.
\par
The Dirac equation, which is of Schr\"odinger type, can of
course describe a c-number fermion system, but the
Klein-Gordon and Maxwell equations, although they are
linear, are not of Schr\"odinger type.  For
example, in one spatial dimension a discretized version
of the Klein-Gordon equation is
\begin{equation}
\ddot q_i -
(1/(2\Delta x))^2
(q_{i + 2} - 2q_i + q_{i - 2}) +
m^2 q_i
= 0.
\label{cx1}\end{equation}
This can be replaced by the first-order equation pair
\begin{equation}
\dot q_i = p_i, \quad \dot p_i =
(1/(2\Delta x))^2
(q_{i + 2} - 2q_i + q_{i - 2}) -
m^2 q_i,
\label{cx2}\end{equation}
which is a version of Hamilton's equations for the
particular Hamiltonian 
(time evolution generating function and observable)
\begin{equation}
H(\vec q,\vec p) = {1\over 2}\sum_k\left (p_k^2 +
(1/(2\Delta x))^2
(q_{k + 1} - q_{k - 1})^2 +
m^2 q_k^2\right ).
\label{cx3}\end{equation}
The constraint equations Eq. (\ref{notes13}) on fermion system
c-number generating functions $G$, which
were previously written in terms of the complex
$(\vec a,\vec a^{\, *})$ vector phase
space variables, translate in terms of the real
$(\vec q,\vec p)$ vector phase space variables
into the pair of real-valued constraint equations:
\begin{equation}
{\partial^2 G\over\partial q_i\partial q_j} =
{\partial^2 G\over\partial p_i\partial p_j}\, , \quad
{\partial^2 G\over\partial q_i\partial p_j} =
-\, {\partial^2 G\over\partial q_j\partial p_i}.
\label{cx4}\end{equation}
For the discretized Klein-Gordon Hamiltonian
given above, we have that
\begin{equation}
{\partial^2 H\over\partial q_i\partial q_{i + 2}} =
-(1/(2\Delta x))^2 \neq 0
\quad\hbox{and}\quad
{\partial^2 H\over\partial p_i\partial p_{i + 2}} = 0,
\label{ cx5}\end{equation}
which is not in accord with the first of the
preceding pair of c-number fermion system
generating function constraint equations.
Thus the Klein-Gordon equation is not of Schr\"odinger
type and cannot describe a c-number fermion
system.
\par
We have seen that c-number fermion dynamics is necessarily described by a 
Schr\"o\-ding\-er type equation, i.e., is necessarily already first quantized, 
and it has no classical version Therefore, its quantization with 
anticommutators is inevitably second quantization. On the other hand, the 
boson commutation relations are consistent with the results of the 
antisymmetric c-number bracket equations, and the first and second quantized 
theories involving bosons commutation relations can be directly related to the 
classical theories through the displacement states Eq. (\ref{o4}). 

\section{Time development of $\lowercase{a_i}$ and $\lowercase{\ah_i}$}

\par
From the infinitesimal transformations which preserve the brackets, it is 
possible to obtain the global representations of the operators which produce 
the time development of the coordinates $a_i$. For the c-number antisymmetric 
bracket case,  the time development operator obtained from Eq. (\ref{it1}) is 
\begin{eqnarray}
a_i(t) &&=U(t)a_i=e^{i t\delta_-(H)}a_i \nonumber\\
\delta_-(H)a_i  &&=\{H,a_i\}_- \nonumber\\
\delta_-^2(H)a_i &&=\{H,\{H,a_i\}_-\}_-, etc. \nonumber\\
a_i(0) &&=a_i. \nonumber\\
\label{td1}\end{eqnarray}
As an example, if $H=a^*a$, then one finds $a(t)=e^{-it}a$.
Under time development, one can show that the antisymmetric bracket is
invariant,
\[ \{a_i(t),a_j^*(t)\}_-=\{a_i,a_j^*\}_-=\delta_{ij}. \]
The proof is as follows:
\begin{eqnarray}
 &&\{a_i(t),a_j^*(t)\}_- \nonumber\\
 &&=\sum_{n=0}^\infty\sum_{m=0}^\infty 
{(it)^{n+m} \over n!m!}\{\delta^n_-(H)a_i,\delta^m_-(H)a_j^*\}_- \nonumber\\
 &&=\sum_{p=0}^\infty
{(it)^{p} \over p!}\sum_{m=0}^p{p! \over (p-m)!m!} 
\{\delta^{p-m}_-(H)a_i,\delta^m_-
(H)a_j^*\}_- \nonumber\\
 &&=\sum_{p=0}^\infty
{(it)^{p}\delta_-^p(H)\over p!}\{a_i,a_j^*\}_-
=e^{i t\delta_-(H)}\{a_i,a_j^*\}_-=\delta_{ij}. \nonumber\\
\label{td2}\end{eqnarray}
In the above, $p=n+m$, and use has been made of 
\[ \delta_-^p(H)\{a_i,a_j^*\}_-
=\sum_{m=0}^p {p \choose m}\{\delta^{p-m}_-(H)a_i,\delta^{m}_-(H)a_j^*\}_-, \]
which follows from the Jacobi identity
\[ \delta_-(H)\{a_i,a_j^*\}_-=\{\delta_-(H)a_i,a_j^*\}_-
+\{a_i,\delta_-(H)a_j^*\}_-. \]
\par The time development generated by the c-number symmetric bracket can be 
studied in a similar manner. Here the time development operator obtained 
from
Eq. (\ref{it5}) for the c-number phase space coordinates is 
\begin{eqnarray}
a_i(t) &&=V(t)a_i=e^{-i t\delta_+(g)}a_i \nonumber\\
\delta_+(g)a_i  &&=\{g,a_i\}_+ \nonumber\\
\delta_+^2(g)a_i &&=\{g,\{g,a_i\}_+\}_+, etc. \nonumber\\
a_i(0) &&=a_i, \nonumber\\
\label{td4}\end{eqnarray}
and $g={\vec a^*}\cdot {\hat G}\cdot {\vec a}$.
Since
\[ \{g,a_i\}_+=\sum_{j}G_{ij}a_j, \]
one finds
\begin{equation}
a_i(t)=\sum_j(\delta_{ij}-itG_{ij}+{(it)^2\over 2!}\sum_k G_{ik}G_{kj}
+\dots) a_{j}.
\label{td4a}\end{equation}
Defining the operator $\hat G$, which must be Hermitian since $g$ is real, 
as done after Eq. (\ref{se2}), we see that
Eq. (\ref{td4a}) becomes                                                           
\begin{eqnarray}
a_i(t)
 &&=\sum_j\bra{i}(I-it{\hat G}+{(it)^2\over 2!}{\hat G}^2+\dots)\ket{j}a_j \nonumber\\
 &&=\bra{i}e^{-it\hat G}\ket{\psi}=\scp{i}{\psi(t)}, \nonumber\\
\label{td5}\end{eqnarray}
since $a_j=\scp{j}{\psi(0)}=\scp{j}{\psi}$.
\par 
It is now easy to demonstrate that the invariance of 
the c-number symmetric bracket
results from the unitary transformation ${\hat U}(t)=e^{-it\hat G}$. 
This is seen from
\begin{eqnarray}
 &&\{a_i(t),a^*_j(t)\}_+
=\{\bra{i}{\hat U}(t)\ket{\psi},\bra{\psi}{\hat U}^\dagger(t)\ket{j}\}_+, \nonumber\\
 &&=\sum_k\sum_l\{\bra{i}{\hat 
U}(t)\ket{k}\bra{k}\psi
\rangle,\langle\psi\ket{l}\bra{l}{\hat U}^\dagger(t)\ket{j}\}_+ \nonumber\\
 &&=\sum_k\sum_l\bra{i}{\hat 
U}(t)\ket{k}\bra{l}{\hat U}^\dagger(t)\ket{j}\{a_k,a_l^*\}_+ \nonumber\\
 &&=\sum_k\sum_l\bra{i}{\hat 
U}(t)\ket{k}\bra{l}{\hat U}^\dagger(t)\ket{j}\delta_{kl}
=\bra{i}{\hat U}(t){\hat U}^\dagger(t)\ket{j}=\delta_{ij}. \nonumber\\
\label{td6}\end{eqnarray}
\par
In the case of the quantum boson or fermion operators, the invariance 
of the commutation relations Eq. (\ref{o2})
follows from the unitarity of the time development operator,
${\hat U}(t)=e^{-it\H}$ obtained from Eq. (\ref{it2}) for the boson case 
or ${\hat U}(t)=e^{-it\hat G}$ obtained from
Eq. (\ref{td5}) for the fermion case, such that
\begin{eqnarray}
\ah_i(t) &&={\hat U}^\dagger(t)\ah_i {\hat U}(t)
=\sum_{n=0}^\infty {(it)^n{\hat D}^n({\hat H})\over n!}\ah_i \nonumber\\
{\hat D}({\hat H})\ah_i  &&=[{\hat H},\ah_i] \nonumber\\
{\hat D}^2({\hat H})\ah_i  &&=[{\hat H},[{\hat H},\ah_i]], etc. \nonumber\\
\ah_i(0) &&=\ah_i. \nonumber\\
\lbrack\ah_i(t),\ad_j(t)\rbrack_{\pm} &&={\hat U}^\dagger(t)\lbrack\ah_i,\ad_j\rbrack_{\pm}{\hat U}(t)
=\delta_{ij}. \nonumber\\
\label{td3}\end{eqnarray}

\section{Angular Momentum}

\par
It is interesting to note that the c-number antisymmetric bracket generates 
the algebra of orbital angular momentum, and that there is a c-number 
differential operator representation of the $SU(2)$ Lie algebra. The 
components of orbital angular momentum are (for $i,j$, and $k=1,2$ or $3$) 
\begin{equation}
l_i=i\sum_i\sum_j\epsilon_{ijk}a_ia_j^*,
\label{am0}\end{equation}
with $\epsilon_{ijk}$ antisymmetric in its indices and $\epsilon_{123}=1$.
For the classical coordinates $a_i{}$, the following relations are found:
\begin{eqnarray}
\{l_i,l_j\}_- &&=i\sum_k\epsilon_{ijk}l_k,\nonumber \\
\{l^2,l_j\}_- &&=0\nonumber \\
\{l_i,q_j\}_- &&
=i\sum_k\epsilon_{ijk}q_k,\,\,\,\{l_i,p_j\}_-=i\sum_k\epsilon_{ijk}p_k.\nonumber \\
\label{am1}\end{eqnarray}
A differential operator representation for the $SU(2)$ algebra is given 
by
\begin{eqnarray}
{\hat J}_{\pm} &&={\hat J}_1\pm i{\hat J}_2   \nonumber\\ 
{\hat J}_+ &&=a^*{\partial \over \partial a},
\,\,\,\, {\hat J}_-=a{\partial \over \partial a^*}   \nonumber\\ 
{\hat J}_3 &&={1\over 2}
(a^*{\partial \over \partial a^*}-a{\partial \over \partial a}) \nonumber\\
\lbrack{\hat J}_3,{\hat J}_{\pm}\rbrack u(j,m) &&=\pm {\hat J}_{\pm}u(j,m)  \nonumber\\
\lbrack{\hat J}_+,{\hat J}_-\rbrack u(j,m) &&=2{\hat J}_3u(j,m)=2m u(j,m)  \nonumber\\
u(j,m) &&={a^{*j+m}a^{j-m}\over\sqrt{(j+m)!(j-m)!}}  \nonumber\\
{\hat J}^2u(j,m) &&={\hat K}({\hat K}+1)u(j,m)=j(j+1)u(j,m)  \nonumber\\
{\hat K} &&={1\over 2}
(a^*{\partial \over \partial a^*}+a{\partial \over \partial a})  \nonumber\\
{\hat J}_{\pm}u(j,m) &&=\sqrt{j(j+1)-m(m\pm 1)}u(j,m\pm 1).  \nonumber\\
\label{am2}\end{eqnarray}
For the functions $u(j,m)$ the inner product is define, with $j\geq j'$, as
\begin{eqnarray}
\langle u(j,m)|u(j',m')\rangle  
&&={1\over 2\pi N(j,m)}\int_0^{2\pi}{\partial^{4j} u^*(j,m)u(j',m')\over
\partial^{2j}a^*\partial^{2j}a}d\phi\nonumber \\
 &&=\delta_{jj'}\delta_{mm'}\nonumber \\
N(j,m) &&={(2j)!(2j)!\over(j+m)!(j-m)!}\nonumber \\
\label{am3}\end{eqnarray}
with $a=\rho e^{i\phi}$.

\section{Conclusions}

\par 
In this paper it has been shown that both quantum theory and the quantum field 
theory of bosons and fermions are a natural consequence of the principle of 
invariance of the symmetric bracket, a concept which is analogous to the 
bracket invariance principle that appears in classical dynamics. Just as the 
invariance of the c-number antisymmetric bracket under a one parameter 
canonical transformation leads to dynamical equations, which determine the 
classical dynamical flow in coordinate phase space, the invariance of the
c-number symmetric bracket under a one parameter canonical transformation leads 
to a dynamical equation, Eq. (\ref{it5}), which determines the  dynamical flow of 
quantum states. In the former case, the dynamical equations are Hamilton's 
equations; however, in the later, the dynamical equation is a time-dependent 
Schr\"odinger type equation Eq. (\ref{notes17}), which is equivalent to generalized 
Hamilton's equations Eq. (\ref{notes11}) or Eq. (\ref{notes12}) for the real part $q_i$ and 
imaginary part $p_i$ of the coordinates $a_i(t)=\scp{i}{\psi(t)}$.  The truly 
remarkable consequences of the principle of invariance of the symmetric 
bracket are the derivation of the time-dependent Schr\"odinger equation and 
the quantum bracket relation $[\hat q,\hat p]=i\hbar$. This argument makes the 
time-dependent Schr\"odinger equation a consequence of bracket invariance, and 
it replaces with a logical derivation the heuristic conjectures of 
Schr\"odinger \cite{esh2},\cite{esh4}, and  \cite{esh5} leading to the 
discovery of his famous equation. Furthermore, it removes $[\hat q,\hat 
p]=i\hbar$ and the time-dependent Schr\"odinger equation  from the status of 
postulates of quantum theory. Along with these results comes the association 
naturally of expectation values of quantum operators with corresponding 
classical quantities, and the statistical interpretation of quantum theory. In 
addition, the c-number time development equation Eq. (\ref{it5}) found from this 
principle provides a natural condition for the emergence of the quantum field 
theory of bosons and fermions, when the antisymmetric bracket is associated 
with the boson operator commutator and the symmetric bracket is associated 
with the fermion operator anticommutator.  It is clear that both the first 
quantized and second quantized theories of bosons have an associated c-number 
dynamics; namely, classical dynamics and classical field theory. These are 
found from the expectation values and matrix elements of operators using boson 
minimum uncertainty displacement states. However, fermion c-number dynamics is 
not classical dynamics, but it is already a first quantized theory, as seen 
from the derivation in Eq. (\ref{se4}). The second quantized version is the quantum 
field theory of fermions. The c-number coordinates in this case satisfy the 
c-number dynamical equation Eq. (\ref{it5}), and they are found as matrix elements 
of fermion operators using the fermion displacement states. 
\par It is clear that the 
fermion dynamics resulting from the symmetric bracket invariance, allows the 
gauge couplings that are known to lead to renormalizable theories for fermion 
dynamics, i.e. QED, and QCD. The old four-fermion theory of beta decay, which 
did not require the intermediation of the W boson, clearly has an equation of 
motion which involves fermion phase space variables in a nonlinear fashion, 
which thus cannot be of the (necessarily linear) Schr\"odinger equation type 
that is here required by invariance of the symmetric bracket. This does not 
mean that effective theories with nonlinear fermion interactions are not 
useful approximations. Examples of such approximate theories are the Hubbard 
model \cite{Frad} and the composite vector boson model \cite{TGPR}, where 
nonlinear interactions may be introduced using path integral methods with 
auxiliary fields. 
\vfill\eject

\end{document}